\documentclass[review]{elsarticle}
\usepackage[T1]{fontenc}
\usepackage{amsmath,amssymb}
\usepackage{booktabs}
\usepackage{multirow}
\usepackage{enumitem}
\usepackage{hyperref}

\journal{Information and Software Technology}

\begin{document}

\begin{frontmatter}

\title{What Aggregate Scores Miss: Measuring Item-Level Regressions in Commercial LLM API Migrations}

\author[gt]{Xiaonan Xu\corref{cor1}}
\ead{xiaonanxu5@gmail.com}
\author[cub]{Wenjing Wu}
\ead{wuwenjing256@gmail.com}
\cortext[cor1]{Corresponding author.}
\affiliation[gt]{organization={College of Computing, Georgia Institute of Technology},
                 city={Atlanta}, state={GA}, postcode={30332}, country={USA}}
\affiliation[cub]{organization={Department of Computer Science, University of Colorado Boulder},
                  city={Boulder}, state={CO}, postcode={80309}, country={USA}}

\begin{abstract}
\textbf{Context:} Software systems that depend on commercial large language model APIs must migrate to successor versions when vendors deprecate older models. Migration decisions typically rely on aggregate benchmark scores, which compress heterogeneous item-level behaviour into a single net figure.
\textbf{Objective:} We measure what that compression conceals.
\textbf{Method:} On three pairwise upgrades in the GPT-5.4 to GPT-5.6~Sol product sequence, we query 900 public benchmark items (graduate-level knowledge, olympiad mathematics, instruction following) 50 times per item per model, classify each item as reliably improved, reliably regressed, practically equivalent, or inconclusive under false-discovery-rate control and a practical-significance threshold, and calibrate the results against a label-permutation null.
\textbf{Results:} Across all nine migration--benchmark cells, reliable improvements and reliable regressions coexist. Edges with aggregate gains of up to 7.3 percentage points contain up to 8.3\% reliably regressed items; edges with aggregate losses contain up to 10.7\% reliably improved items. On the instruction-following benchmark, the gap between strict and loose scoring widens by 3.9 percentage points on the latest migration: a 3.9-point regression under strict scoring shrinks to 0.04 points under loose scoring.
\textbf{Conclusion:} Migration decisions based on aggregate scores alone miss substantial bidirectional item-level change. The complete response-level archive and per-item scoring outputs are released.
\end{abstract}

\begin{keyword}
LLM evaluation \sep backward compatibility \sep regression testing \sep API migration \sep repeated sampling
\end{keyword}

\end{frontmatter}

\section{Introduction}
\label{sec:introduction}

Commercial large language model (LLM) APIs have become external software dependencies in a growing share of production systems~\cite{hou2024llm4se}. These APIs are controlled by the vendor: models are retired on published deprecation schedules, and downstream consumers must migrate to successor versions or lose access~\cite{ma2024prompt}. Library-migration studies show that forced dependency updates carry real cost for client projects~\cite{kula2018migration}, and that updates preserving interfaces can still break client behaviour~\cite{jayasuriya2024behavioral}. Each migration replaces the model behind every downstream call site. Whether the new version preserves the test-case-level behaviour of the old one, a property known in software engineering as backward compatibility, determines whether the migration is safe.

Standard benchmark reporting compresses the answer to that question into aggregate scores. A net gain of two percentage points on a benchmark may consist of 100 items answered more reliably and 80 answered less reliably; the aggregate reports only the net balance. This compression has practical consequences: in July 2026, OpenAI's Thibault Sottiaux publicly stated that, before the GPT-5.6 Sol launch, the team had focused on average and median usage and missed cases in which long-tail usage was substantially higher~\cite{sottiaux2026}. The stochastic nature of LLM outputs adds a second difficulty. A single correct-to-incorrect flip between two model versions may reflect sampling noise alone. Ma et al.~\cite{ma2024prompt} identify non-determinism as one of three fundamental obstacles to applying regression testing to LLM APIs, and in empirical measurements on open-weight models, single-draw evaluation missed 42\% of reliably changed items~\cite{beyond_the_mean_2026}. Backward compatibility at the item level therefore requires estimating pass-probability changes and calibrating them against a permutation null.

A longitudinal study of 18 GPT models fits ability trajectories to single-draw responses, so estimated probability shifts share one direction within each comparison~\cite{paper8871_2026}. A repeated-sampling study on open-weight 7--8B models measures bidirectional reliable churn at $K{=}10$, leaving frontier commercial models as an open question~\cite{beyond_the_mean_2026}. An industry evaluation compares GPT-5.5 and GPT-5.6 on 711 enterprise workflows at 5 runs per task with Holm-corrected significance tests~\cite{toloka2026}. Our study extends this repeated-sampling approach to frontier commercial API migrations at $K{=}50$ with permutation-null calibration of the complete classification procedure.

We compare three pairwise upgrades in the GPT-5.4 to GPT-5.6~Sol product sequence on 900 public benchmark items spanning graduate-level knowledge, olympiad mathematics, and instruction following, with all request parameters held constant except the model identifier. Each model--item pass probability is estimated from 50 independent trials, and item-level judgements are calibrated against a permutation null at matched sample size. On the instruction-following benchmark we also compare the official strict and loose verifier scores to test whether measured regressions depend on exact output compliance. We find that reliable improvements and reliable regressions coexist in all nine migration--benchmark cells: edges with a positive aggregate change contain up to 8.3\% reliably regressed items, and edges with a negative aggregate change contain up to 10.7\% reliably improved items. On the instruction-following benchmark, the strict--loose scoring gap widens by 3.9 percentage points on the latest migration: the same edge shows a 3.9-point regression under strict scoring and a 0.04-point regression under loose scoring.

The contributions of this paper are as follows.
\begin{enumerate}[leftmargin=*, nosep]
\item We measure item-level backward compatibility across the GPT-5.4 to GPT-5.6~Sol product sequence, querying 900 public benchmark items 50 times per item per model. Item-level judgements are calibrated against a permutation null that reruns the complete classification procedure at matched sample size.
\item We show that reliable improvements and reliable regressions coexist in all nine migration--benchmark cells, including edges where the aggregate score improved.
\item We quantify how migration conclusions change between strict and loose scoring: on the instruction-following benchmark, the latest migration widens the gap between strict and loose scoring, causing the regression observed under strict scoring to shrink under loose scoring.
\item We release the complete response-level archive with per-item scoring outputs, enabling verification and alternative rescoring without re-querying mutable commercial APIs.
\end{enumerate}

\section{Related Work}
\label{sec:related}

\paragraph{Aggregate evaluation and its limits}
The dominant practice in LLM evaluation is to report aggregate scores on static benchmarks. Chang et al.~\cite{chang2024survey} survey evaluation practices and identify recurring concerns: sensitivity to prompt format, inconsistency across evaluation pipelines, and the limited diagnostic value of a single aggregate figure. Benchmark saturation and data contamination amplify these concerns as models and training corpora co-evolve~\cite{chen2025contamination}, and a mapping of generative-AI metrics to software quality characteristics concludes that no single aggregate captures the relevant quality dimensions~\cite{yu2025metrics}. Aggregate scores remain useful summaries of overall capability, but they reduce heterogeneous item-level behaviour to a single net figure that conflates uniform improvement with a mixture of gains and losses. Our study takes this compression as its object of measurement and quantifies what the aggregate hides.

\paragraph{Model-update regression in LLM APIs}
Classical regression testing assumes deterministic test outcomes, an assumption already strained within conventional software by flaky tests~\cite{luo2014flaky}; stochastic LLM outputs invalidate it entirely. Chen et al.~\cite{chen2023chatgpt} provided early large-scale evidence that commercial GPT behaviour shifts across time-stamped snapshots, comparing two snapshots at the task level. Ma et al.~\cite{ma2024prompt} reframed the phenomenon as a software engineering problem, identifying three properties of commercial LLM APIs that break classical regression-testing assumptions: vendor-controlled updates, prompt sensitivity, and non-deterministic outputs. Echterhoff et al.~\cite{echterhoff2024muscle} formalised negative flips (items correct before an update and incorrect after) and proposed compatibility-aware training to reduce them, addressing the problem from the model developer's side. A production-oriented migration framework~\cite{migration2026} uses Bayesian candidate comparison to support replacement decisions when a model reaches end-of-life. Dong et al.~\cite{dong2026reliability} showed that instruction-following accuracy on GPT-4o snapshots drops when prompts are paraphrased, decomposing aggregate scores along the prompt-variation axis. Our work shares the decomposition goal but varies the model version rather than the prompt, and measures pass-probability changes from repeated sampling rather than from prompt variants.

\paragraph{Item-level measurement and repeated sampling}
Three recent studies address item-level version comparison directly. A longitudinal study covering 18 GPT models from GPT-3.5 to GPT-5.2~\cite{paper8871_2026} fits a dynamic item response theory model to single-draw binary responses, estimating latent ability trajectories and localising probability changes across difficulty and discrimination regions. Because the model assigns a single ability parameter per snapshot, all item-level probability shifts share the same sign within a comparison; the authors note that heterogeneous directions require analysis of observed response flips, which the single-draw design does not support. The repeated-sampling approach of~\cite{beyond_the_mean_2026} addresses this directly: each item is queried $K{=}10$ times, and a reliable-change index with a permutation null separates true changes from sampling noise on open-weight Llama and Qwen models (7--8B parameters). That study reports 21--28\% total reliable churn; single-draw evaluation missed 42\% of reliably changed items and falsely flagged 25\% of unchanged items. It identifies frontier commercial models and additional benchmarks as open questions. Toloka~\cite{toloka2026} evaluates GPT-5.5 against GPT-5.6 on 711 frozen enterprise workflows with 5 runs per task, applying stratified bootstrap and Holm-corrected significance tests to detect aggregate regressions and identify strong per-task flips. The present study extends the repeated-sampling line to frontier commercial API migrations, raises the per-item trial count to $K{=}50$, applies false-discovery-rate control with a practical-significance threshold, and calibrates item-level judgements against a permutation null.

\paragraph{Output variability}
Repeated queries to the same commercial model return varying answers. Ouyang et al.~\cite{ouyang2025nondeterminism} measure this directly for ChatGPT code generation, finding semantic variation across identical requests even at temperature zero, and Kim and Ming~\cite{kim2025reliability} compare output reliability and similarity across models on software-development tasks. A systematic catalogue attributes such divergence to sampling, silent updates, numerical rounding, and expert routing~\cite{randomness2026}. Estimating per-item pass probabilities from multiple generations, rather than judging single draws, follows the direction formalised by Zhang et al.~\cite{zhang2026multiple}. Community guidelines for LLM-based empirical studies likewise identify output non-determinism and model evolution as threats to reproducibility and call for archived interaction traces~\cite{llmguidelines2025}, a practice the released response archive follows.

\section{Method}
\label{sec:method}

\subsection{Research questions}
\label{sec:rq}

We study whether aggregate benchmark scores reliably reflect the backward compatibility of commercial LLM API upgrades. Three research questions structure the study.

\textbf{RQ1.} On each migration edge in the current GPT product line, what proportion of benchmark items shows reliable improvement, and what proportion shows reliable regression, beyond what sampling noise alone would produce?

\textbf{RQ2.} How does the aggregate score change on each edge relate to the underlying item-level changes, and how much bidirectional change does the aggregate figure conceal?

\textbf{RQ3.} How do migration conclusions change between strict and loose scoring of the same responses?

\subsection{Study design}
\label{sec:design}

The unit of comparison is a \emph{migration edge}: an ordered pair of models that an API consumer moves between when following the vendor's product line. We measure three edges over three models, GPT-5.4, GPT-5.5, and GPT-5.6 Sol: the two consecutive flagship migrations (5.4$\rightarrow$5.5 and 5.5$\rightarrow$Sol) and the direct migration 5.4$\rightarrow$Sol, which the vendor's guidance also supports. The three models are served under the API identifiers \texttt{gpt-5.4}, \texttt{gpt-5.5}, and \texttt{gpt-5.6-sol}.

All request parameters are held constant across models; only the model identifier varies. This mirrors a dependency upgrade in which client code is unchanged and only the dependency version moves. Reasoning effort is set explicitly to \texttt{medium} for every model, the shared tier across the three models' effort ranges. Temperature is not accepted by these models. No output-length cap is imposed. Each item is queried $K = 50$ times per model in independent single-turn calls.

\subsection{Benchmarks and item selection}
\label{sec:benchmarks}

Measuring item-level change requires items on which the models under study retain headroom: an item answered correctly, or incorrectly, in every trial by every model carries no information about migration-induced change. We therefore selected benchmarks by a screening procedure defined before the main collection. Eleven public, automatically scorable candidates spanning knowledge, mathematics, and instruction following were screened with 30 randomly drawn items each, three models, and $K = 10$ repetitions per item. For each candidate we computed the fraction of items on which at least one model's observed accuracy fell in $[0.1,\,0.9]$. One benchmark per category was selected, requiring that at least 35\% of items meet this criterion, and taking the candidate with the highest proportion per slot. Widely used benchmarks, including GPQA Diamond, MMLU-Pro, and IFEval, fell below the threshold on these models, with proportions between 7\% and 27\%; the full screening table is reported in the appendix. Screening data served benchmark selection only and did not enter the main analysis.

The selected benchmarks are SuperGPQA~\cite{supergpqa2025} (knowledge; 500 items drawn at random from the full pool before collection), Omni-MATH hard~\cite{omnimath2024} (mathematics; 100 items drawn at random from the 452-item pool before collection), and IFBench~\cite{ifbench2025} (instruction following; all 300 prompts, with the prompt as the unit of analysis), for a total of 900 items. The sampled item lists were fixed before collection and are included in the released archive.

Prompting follows the benchmark type. Knowledge and mathematics items use a minimal task statement and a final-answer-line convention, frozen after piloting. IFBench prompts are used verbatim from the official release, since their wording constitutes the task constraints; they are scored by the official verifier~\cite{ifbench2025,ifeval2023}.

\subsection{Data collection}
\label{sec:collection}

Collection proceeded in per-model phases in release order (GPT-5.4, then GPT-5.5, then GPT-5.6 Sol), with item--repetition order randomised within each phase. Collection ran from 30 July to 13 August 2026. All phases ran at a fixed request concurrency of 36. Transient API errors were retried with exponential backoff until success; the final response matrix is complete.

\subsection{Scoring}
\label{sec:scoring}

Scoring is decoupled from collection and deterministic given the archived responses. Every response is assigned to one of six mutually exclusive categories: semantically correct, semantically wrong, format failure, refusal, truncation, and API error; given the retry policy, every response in the final matrix falls in the first five categories. API errors are excluded from the accuracy denominator and their occurrence rate is reported separately; refusals, format failures, and model-side truncations remain in the denominator as behavioural failures.

For knowledge and mathematics items, a response is scored correct when a two-stage parser extracts an answer from the final answer line and that answer matches the reference; mathematical equivalence is established by exact match followed by symbolic normalisation. IFBench responses are scored by the official verifiers~\cite{ifbench2025}, which provide a strict and a loose reading; the loose reading re-applies each verification function after a fixed set of surface-format normalisations, so the two readings differ only in format tolerance. The strict reading is primary. Responses that a downstream system could not parse under the agreed format are counted as failures, since for an API consumer an unparseable reply is operationally indistinguishable from a wrong one.

\subsection{Analysis}
\label{sec:analysis}

For item $i$ under model $m$, the observed accuracy is $\hat{p}_{m,i}$, the fraction of the $K$ trials scored correct under the primary (strict) rule. For each edge, source $s$ to target $t$, we report per benchmark the aggregate change $\Delta = \operatorname{mean}_i(\hat{p}_{t,i}) - \operatorname{mean}_i(\hat{p}_{s,i})$; the reliable-improvement share $P^{+}$; the reliable-regression share $P^{-}$; and the reliable-change share $P^{\mathit{chg}} = P^{+} + P^{-}$.

Item-level judgements require both statistical significance and a minimum effect size. A Fisher exact test on the $2 \times K$ correct/incorrect counts, with Benjamini--Hochberg control of the false discovery rate at 5\% within each migration--benchmark cell, establishes significance. A practical-significance threshold $\lvert \hat{p}_{t,i} - \hat{p}_{s,i} \rvert \geq \varepsilon$ with $\varepsilon = 0.2$ (a 20-percentage-point shift in pass probability) establishes a minimum effect size. Items meeting both criteria are classified as reliably improved or reliably regressed by sign. Items whose 95\% confidence interval for the difference lies entirely within $(-\varepsilon, \varepsilon)$ are classified as practically equivalent; the remainder are classified as inconclusive.

Observed shares are calibrated against a permutation null. For each item we pool the $2K$ outcomes of the two models on an edge, randomly reassign version labels with $K$ outcomes per side, and rerun the complete classification procedure; 1{,}000 replications yield the null distribution of $P^{+}$, $P^{-}$, and $P^{\mathit{chg}}$ under zero true change at matched sample size. Reported shares are presented alongside the 95th percentile of this null, so that reliable change is claimed only where it exceeds what label noise alone produces.

For the secondary analysis on IFBench, let $R_m$ denote the mean difference between the loose and strict verifier scores across prompts. Each edge reports $\Delta R = R_t - R_s$, the change in this gap across the migration.

\section{Results}
\label{sec:results}

\subsection{Aggregate scores and item-level changes}
\label{sec:main-results}

\begin{figure*}[t]
\centering
\includegraphics[width=0.95\textwidth]{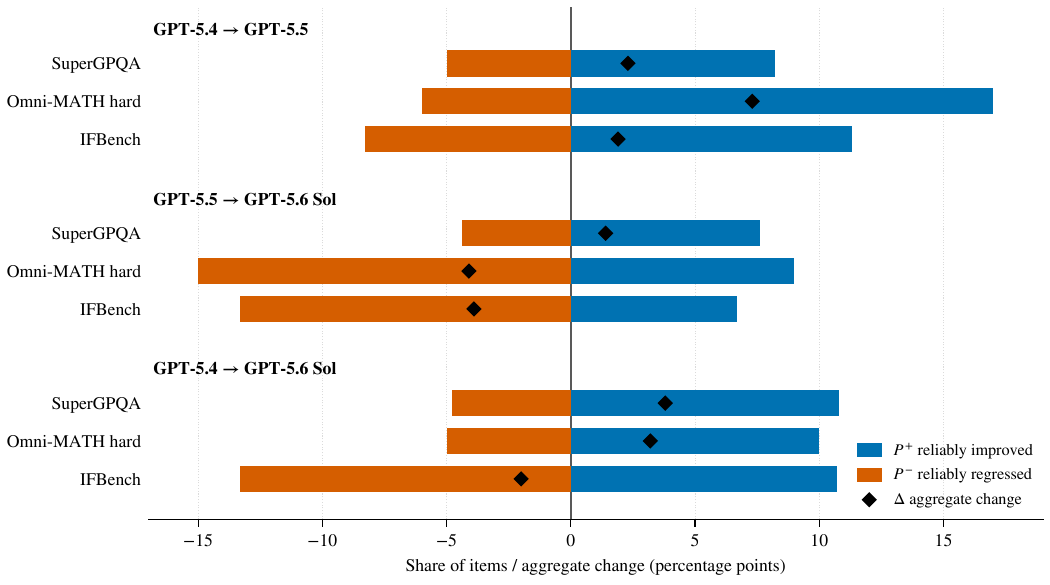}
\caption{Item-level reliable improvement and regression alongside aggregate change for each migration edge and benchmark. Bars extending right (left) show the share of items reliably improved (regressed); the central marker shows the aggregate change~$\Delta$}
\label{fig:main}
\end{figure*}

Mean strict accuracy was 63.4\%, 45.7\%, and 62.8\% for GPT-5.4 on SuperGPQA, Omni-MATH hard, and IFBench; 65.7\%, 53.0\%, and 64.6\% for GPT-5.5; and 67.1\%, 48.9\%, and 60.7\% for Sol. Table~\ref{tab:main} summarises the aggregate change and the full item-level classification for each migration edge and benchmark; Figure~\ref{fig:main} shows the reliable-improvement and reliable-regression shares alongside the aggregate change.

\begin{table*}[t]
\centering
\caption{Aggregate change and item-level classification for each migration edge and benchmark. $\Delta$~is the change in mean strict accuracy in percentage points. $P^{+}$, $P^{-}$, and $P^{\mathit{chg}}$ are the shares of items classified as reliably improved, reliably regressed, and reliably changed under the Fisher/BH/$\varepsilon{=}0.2$ criteria; Equiv.\ and Incon.\ are the practically-equivalent and inconclusive shares. Null$_{95}$~is the 95th percentile of $P^{\mathit{chg}}$ under the permutation null (1{,}000 replications). All shares in percent; $\Delta$~is computed before rounding}
\label{tab:main}
\small
\begin{tabular}{ll rrrrrrr}
\toprule
Edge & Benchmark & $\Delta$ & $P^{+}$ & $P^{-}$ & $P^{\mathit{chg}}$ & Equiv. & Incon. & Null$_{95}$ \\
\midrule
\multirow{3}{*}{5.4$\to$5.5}
 & SuperGPQA      & $+$2.3 & 8.2  & 5.0  & 13.2 & 77.0 & 9.8  & 0.0 \\
 & Omni-MATH hard & $+$7.3 & 17.0 & 6.0  & 23.0 & 59.0 & 18.0 & 0.0 \\
 & IFBench        & $+$1.9 & 11.3 & 8.3  & 19.7 & 55.7 & 24.7 & 0.0 \\
\midrule
\multirow{3}{*}{5.5$\to$Sol}
 & SuperGPQA      & $+$1.4 & 7.6  & 4.4  & 12.0 & 80.4 & 7.6  & 0.0 \\
 & Omni-MATH hard & $-$4.1 & 9.0  & 15.0 & 24.0 & 67.0 & 9.0  & 0.0 \\
 & IFBench        & $-$3.9 & 6.7  & 13.3 & 20.0 & 57.0 & 23.0 & 0.0 \\
\midrule
\multirow{3}{*}{5.4$\to$Sol}
 & SuperGPQA      & $+$3.8 & 10.8 & 4.8  & 15.6 & 76.8 & 7.6  & 0.0 \\
 & Omni-MATH hard & $+$3.2 & 10.0 & 5.0  & 15.0 & 70.0 & 15.0 & 0.0 \\
 & IFBench        & $-$2.0 & 10.7 & 13.3 & 24.0 & 54.3 & 21.7 & 0.0 \\
\bottomrule
\end{tabular}
\end{table*}

On the 5.4$\to$5.5 edge, aggregate scores rose on all three benchmarks ({+2.3}, {+7.3}, and {+1.9} percentage points on SuperGPQA, Omni-MATH hard, and IFBench). All three benchmarks also show reliable regression: 5.0\% of SuperGPQA items, 6.0\% of Omni-MATH items, and 8.3\% of IFBench items regressed reliably despite a positive aggregate change.

On the 5.5$\to$Sol edge, the three benchmarks diverge. SuperGPQA continued to rise ({+1.4}) while Omni-MATH hard ({$-$4.1}) and IFBench ({$-$3.9}) declined. Both declining benchmarks still contain reliably improved items: 9.0\% of Omni-MATH items and 6.7\% of IFBench items improved reliably even as the aggregate fell.

On the direct 5.4$\to$Sol edge, SuperGPQA showed the largest gain ({+3.8}) and the highest improvement share (10.8\%), while IFBench showed a net loss ({$-$2.0}) with 13.3\% of items reliably regressed and 10.7\% reliably improved. Reliable-change shares on this edge differ from the sums of the two consecutive edges, reflecting item-level movements that cancel across consecutive migrations.

\subsection{Permutation-null calibration}
\label{sec:null}

Under zero true change, the 95th percentile of the reliable-change share was 0.0\% in every migration--benchmark cell: across 1{,}000 label permutations at the study's sample sizes, the Fisher/BH/$\varepsilon$ criteria produced zero spurious reliable changes. Every reliably changed item in Table~\ref{tab:main} therefore lies strictly above the noise floor. Prior work on open-weight 7--8B models reported total churn of 21--28\% at $K{=}10$~\cite{beyond_the_mean_2026}; the shares of 12.0--24.0\% observed here are of comparable magnitude yet rest on a calibrated zero baseline.

\subsection{Strict--loose scoring gap on IFBench}
\label{sec:contract}

IFBench's official verifiers provide a strict and a loose reading of every response. The gap $R_m$ between the two readings was 6.3 percentage points for GPT-5.4, 5.1 for GPT-5.5, and 9.0 for Sol, yielding a $\Delta R$ of $-$1.2 points on the 5.4$\to$5.5 edge, {+3.9} on 5.5$\to$Sol, and {+2.7} on 5.4$\to$Sol. The widening on the 5.5$\to$Sol edge accounts for nearly the entire aggregate decline: Sol's strict score falls 3.9 points below GPT-5.5's, while its loose score falls 0.04 points; the regression is concentrated in exact constraint compliance. On SuperGPQA and Omni-MATH hard, responses are scored by answer extraction and exact match; extraction failures were negligible, and the six-category breakdown in the appendix reports their rates.

\subsection{Supplementary analyses}
\label{sec:appendix-preview}

The appendix reports five supplementary analyses on the same data: a single-draw comparison that contrasts raw correct-to-incorrect and incorrect-to-correct flips from one randomly selected trial per item with the $K{=}50$ classification; a sensitivity analysis of the practical-significance threshold over $\varepsilon \in \{0.10, 0.15, 0.20, 0.25\}$; token-cost distributions per model and edge; the six-category response breakdown; and difficulty stratification using metadata provided by each benchmark.

\section{Discussion}
\label{sec:discussion}

Across all nine migration--benchmark cells, reliable improvements and reliable regressions occur together. The permutation null confirms that the observed shares exceed what sampling noise alone produces at $K{=}50$. Aggregate scores therefore provide a necessary but insufficient basis for migration decisions: on the six cells where the aggregate improved, a gate based on the aggregate alone would have accepted the migration while 4.4--8.3\% of items regressed reliably.

The IFBench results raise a separate concern. Because the two readings differ only in format tolerance, the widening of the strict--loose gap on the 5.5-to-Sol edge locates part of the observed strict-scoring regression in exact format compliance rather than in the underlying task. For a system that parses outputs programmatically, format non-compliance is a functional regression; a consumer that tolerates format variation faces a smaller compatibility cost from the same migration. Migration risk therefore depends on the acceptance criteria applied to model outputs, and capturing both perspectives requires reporting both readings.

The repeated-sampling study on open-weight models~\cite{beyond_the_mean_2026} asked whether its findings extend to frontier commercial models; the present measurements provide a direct comparison point at $K{=}50$ on three GPT versions. The Toloka evaluation~\cite{toloka2026} identified per-task regressions on the same product line at $K{=}5$ with Holm-corrected tests; our higher per-item trial count and permutation null complement that evidence by separating reliable item-level changes from sampling artefacts at higher statistical resolution.

The appendix compares single-draw evaluation with the repeated-sampling judgements directly: a single binary observation per model cannot reach significance under the item-level test, so none of the 457 reliable item-level changes across the nine cells is detectable from one draw, and the appendix additionally reports how raw single-draw flips distribute over the $K{=}50$ classes. The gap between single-draw and $K{=}50$ judgements sets a lower bound on the information lost when migration decisions rely on single-draw evaluation. Organisations that treat model upgrades as dependency updates can use the item-level regression shares reported here as reference rates when designing acceptance tests for their own workloads.

\section{Threats to Validity}
\label{sec:threats}

Three limitations bound the scope of these measurements. The benchmark screening step selects benchmarks on which the studied models retain headroom; the reported reliable-change shares are therefore estimates for informative benchmarks, not population rates over arbitrary workloads. The study covers one vendor's product line (GPT-5.4 through GPT-5.6~Sol) and three public benchmarks spanning knowledge, mathematics, and instruction following; other vendors, task types, and production workloads may show different patterns. The practical-significance threshold $\varepsilon = 0.2$ and the trial count $K{=}50$ together determine which item-level changes are detectable: smaller genuine changes remain in the inconclusive category and are reported as such.

\section{Conclusion}
\label{sec:conclusion}

Across three pairwise upgrades in the GPT-5.4 to GPT-5.6~Sol product sequence, with 900 public benchmark items, 50 trials per item, and calibration against a permutation null, every migration--benchmark cell contains both reliably improved and reliably regressed items; aggregate gains of up to 7.3 percentage points accompany up to 8.3\% reliably regressed items. The strict--loose scoring gap widens on the latest migration for instruction following, indicating that part of the measured regression concentrates on format compliance. The complete response-level archive and per-item scoring outputs are released to support verification and alternative analyses.

\section*{Data Availability}

The complete response-level archive (request parameters, raw responses, returned model identifiers, usage metadata, and timestamps) and the per-item scoring outputs are available at \url{https://github.com/WenJing95/gpt-regression-data}. All reported statistics follow the procedures specified in Section~\ref{sec:method} and can be recomputed from this archive without re-querying commercial APIs.

\section*{Funding}

This research did not receive any specific grant from funding agencies in the public, commercial, or not-for-profit sectors.

\section*{Declaration of Competing Interest}

The authors declare that they have no known competing financial interests or personal relationships that could have appeared to influence the work reported in this paper.

\section*{Declaration of generative AI and AI-assisted technologies in the writing process}

During the preparation of this work the authors used Claude (Anthropic) and ChatGPT (OpenAI) in order to improve the language and readability of the manuscript. After using these tools, the authors reviewed and edited the content as needed and take full responsibility for the content of the published article.

\section*{CRediT authorship contribution statement}

\textbf{Xiaonan Xu:} Conceptualization, Methodology, Investigation, Data curation, Writing -- original draft. \textbf{Wenjing Wu:} Software, Validation, Formal analysis, Writing -- review \& editing.

\bibliographystyle{elsarticle-num}
\bibliography{references}

@article{chang2024survey,
  author    = {Yupeng Chang and Xu Wang and Jindong Wang and Yuan Wu and Linyi Yang and Kaijie Zhu and Hao Chen and Xiaoyuan Yi and Cunxiang Wang and Yidong Wang and others},
  title     = {A Survey on Evaluation of Large Language Models},
  journal   = {ACM Transactions on Intelligent Systems and Technology},
  volume    = {15},
  number    = {3},
  pages     = {39:1--39:45},
  year      = {2024},
}

@inproceedings{chen2025contamination,
  author    = {Simin Chen and Yiming Chen and Zexin Li and Yifan Jiang and Zhongwei Wan and Yixin He and Dezhi Ran and Tianle Gu and Haizhou Li and Tao Xie and Baishakhi Ray},
  title     = {Benchmarking Large Language Models Under Data Contamination: A Survey from Static to Dynamic Evaluation},
  booktitle = {Proceedings of the 2025 Conference on Empirical Methods in Natural Language Processing},
  pages     = {10080--10098},
  doi       = {10.18653/v1/2025.emnlp-main.511},
  year      = {2025},
}

@article{hou2024llm4se,
  author    = {Xinyi Hou and Yanjie Zhao and Yue Liu and Zhou Yang and Kailong Wang and Li Li and Xiapu Luo and David Lo and John Grundy and Haoyu Wang},
  title     = {Large Language Models for Software Engineering: A Systematic Literature Review},
  journal   = {ACM Transactions on Software Engineering and Methodology},
  volume    = {33},
  number    = {8},
  pages     = {220:1--220:79},
  year      = {2024},
}

@inproceedings{ma2024prompt,
  author    = {Wanqin Ma and Chenyang Yang and Christian K\"{a}stner},
  title     = {({Why}) Is My Prompt Getting Worse? {R}ethinking Regression Testing for Evolving {LLM} {API}s},
  booktitle = {Proceedings of the IEEE/ACM 3rd International Conference on AI Engineering -- Software Engineering for AI (CAIN)},
  pages     = {166--171},
  year      = {2024},
}

@article{chen2023chatgpt,
  author    = {Lingjiao Chen and Matei Zaharia and James Zou},
  title     = {How Is {ChatGPT}'s Behavior Changing over Time?},
  journal   = {arXiv preprint arXiv:2307.09009},
  year      = {2023},
}

@inproceedings{echterhoff2024muscle,
  author    = {Jessica Maria Echterhoff and Fartash Faghri and Raviteja Vemulapalli and Ting-Yao Hu and Chun-Liang Li and Oncel Tuzel and Hadi Pouransari},
  title     = {{MUSCLE}: A Model Update Strategy for Compatible {LLM} Evolution},
  booktitle = {Findings of the Association for Computational Linguistics: EMNLP 2024},
  pages     = {7320--7332},
  year      = {2024},
}

@article{beyond_the_mean_2026,
  author    = {Jon-Paul Cacioli},
  title     = {Beyond the Mean: Within-Model Reliable Change Detection for {LLM} Evaluation},
  journal   = {arXiv preprint arXiv:2604.27405},
  year      = {2026},
}

@unpublished{paper8871_2026,
  author    = {Anonymous},
  title     = {Longitudinal Evaluation of Large Language Models},
  note      = {Under review for Transactions on Machine Learning Research (TMLR), Paper 8871. Submitted 2026-05-11. \url{https://openreview.net/forum?id=INuSvLC7Bq}},
  year      = {2026},
}

@misc{toloka2026,
  author    = {{Toloka AI}},
  title     = {{GPT-5.6} Got Smarter. {T}hen It Kept Acting.},
  howpublished = {\url{https://toloka.ai/blog/gpt-5.6-got-smarter-then-it-kept-acting/}},
  note      = {Blog post, July 2026. Accessed 16 August 2026},
  year      = {2026},
}

@article{randomness2026,
  author    = {Guillaume Coqueret and Joan Llull and Florian Oswald and Christophe P\'{e}rignon and Christoph Scheuch and Lars Vilhuber},
  title     = {Randomness in Large Language Models: What Researchers Need to Know (and Report)},
  journal   = {arXiv preprint arXiv:2607.24372},
  year      = {2026},
}

@article{migration2026,
  author    = {Emma Casey and David Roberts and David Sim and Ian Beaver},
  title     = {When Your {LLM} Reaches End-of-Life: A Framework for Confident Model Migration in Production Systems},
  journal   = {arXiv preprint arXiv:2604.27082},
  year      = {2026},
}

@inproceedings{dong2026reliability,
  author    = {Jianshuo Dong and Yutong Zhang and Yan Liu and Zhenyu Zhong and Tao Wei and Chao Zhang and Han Qiu},
  title     = {Revisiting the Reliability of Language Models in Instruction-Following},
  booktitle = {Proceedings of the 64th Annual Meeting of the Association for Computational Linguistics (Volume 1: Long Papers)},
  pages     = {7784--7812},
  doi       = {10.18653/v1/2026.acl-long.354},
  year      = {2026},
}

@article{supergpqa2025,
  author    = {M-A-P Team},
  title     = {{SuperGPQA}: Scaling {LLM} Evaluation across 285 Graduate Disciplines},
  journal   = {arXiv preprint arXiv:2502.14739},
  year      = {2025},
}

@article{omnimath2024,
  author    = {Bofei Gao and Feifan Song and Zhe Yang and Zefan Cai and Yibo Miao and Qingxiu Dong and Lei Li and Chenghao Ma and Liang Chen and Runxin Xu and others},
  title     = {{Omni-MATH}: A Universal Olympiad Level Mathematic Benchmark for Large Language Models},
  journal   = {arXiv preprint arXiv:2410.07985},
  year      = {2024},
}

@inproceedings{ifbench2025,
  author    = {Valentina Pyatkin and Saumya Malik and Victoria Graf and Hamish Ivison and Shengyi Huang and Pradeep Dasigi and Nathan Lambert and Hannaneh Hajishirzi},
  title     = {Generalizing Verifiable Instruction Following},
  booktitle = {Advances in Neural Information Processing Systems 38 (NeurIPS 2025) Datasets and Benchmarks Track},
  year      = {2025},
}

@article{ifeval2023,
  author    = {Jeffrey Zhou and Tianjian Lu and Swaroop Mishra and Siddhartha Brahma and Sujoy Basu and Yi Luan and Denny Zhou and Le Hou},
  title     = {Instruction-Following Evaluation for Large Language Models},
  journal   = {arXiv preprint arXiv:2311.07911},
  year      = {2023},
}

@inproceedings{luo2014flaky,
  author    = {Qingzhou Luo and Farah Hariri and Lamyaa Eloussi and Darko Marinov},
  title     = {An Empirical Analysis of Flaky Tests},
  booktitle = {Proceedings of the 22nd ACM SIGSOFT International Symposium on Foundations of Software Engineering (FSE)},
  pages     = {643--653},
  year      = {2014},
}

@article{llmguidelines2025,
  author    = {Sebastian Baltes and Florian Angermeir and Chetan Arora and Marvin Mu{\~n}oz Bar{\'o}n and Chunyang Chen and Lukas B{\"o}hme and Fabio Calefato and Neil Ernst and Davide Falessi and Brian Fitzgerald and others},
  title     = {Evaluation Guidelines for Empirical Studies in Software Engineering involving {LLMs}},
  journal   = {arXiv preprint arXiv:2508.15503},
  year      = {2025},
}

@article{kula2018migration,
  author  = {Raula Gaikovina Kula and Daniel M. German and Ali Ouni and Takashi Ishio and Katsuro Inoue},
  title   = {Do Developers Update Their Library Dependencies? An Empirical Study on the Impact of Security Advisories on Library Migration},
  journal = {Empirical Software Engineering},
  volume  = {23}, number = {1}, pages = {384--417}, year = {2018},
}

@article{jayasuriya2024behavioral,
  author  = {Dhanushka Jayasuriya and Valerio Terragni and Jens Dietrich and Kelly Blincoe},
  title   = {Understanding the Impact of {API}s Behavioral Breaking Changes on Client Applications},
  journal = {Proceedings of the ACM on Software Engineering},
  volume  = {1}, number = {FSE}, pages = {1238--1261}, year = {2024},
}

@article{ouyang2025nondeterminism,
  author  = {Shuyin Ouyang and Jie M. Zhang and Mark Harman and Meng Wang},
  title   = {An Empirical Study of the Non-Determinism of {ChatGPT} in Code Generation},
  journal = {ACM Transactions on Software Engineering and Methodology},
  volume  = {34}, number = {2}, pages = {42:1--42:28}, year = {2025},
}

@inproceedings{zhang2026multiple,
  author  = {Wenbo Zhang and Hengrui Cai and Wenyu Chen},
  title   = {Beyond the Singular: Revealing the Value of Multiple Generations in Benchmark Evaluation},
  booktitle = {Findings of the Association for Computational Linguistics: ACL 2026},
  pages   = {10033--10043}, year = {2026},
}

@article{yu2025metrics,
  author  = {Liang Yu and Emil Al{\'e}groth and Panagiota Chatzipetrou and Tony Gorschek},
  title   = {Measuring the Quality of Generative {AI} Systems: Mapping Metrics to Quality Characteristics---Snowballing Literature Review},
  journal = {Information and Software Technology},
  volume  = {186}, pages = {107802}, year = {2025},
}

@article{kim2025reliability,
  author  = {Dae-Kyoo Kim and Hua Ming},
  title   = {Assessing Output Reliability and Similarity of Large Language Models in Software Development: A Comparative Case Study Approach},
  journal = {Information and Software Technology},
  volume  = {185}, pages = {107787}, year = {2025},
}

@misc{sottiaux2026,
  author       = {Thibault Sottiaux},
  title        = {Sol community update: {GPT-5.6 Sol} usage quotas},
  howpublished = {X (formerly Twitter), \url{https://x.com/thsottiaux/status/2082317452755751098}},
  note         = {Post of 29 July 2026. Accessed 16 August 2026},
  year         = {2026},
}

\appendix

\section{Benchmark screening}
\label{app:screening}

Table~\ref{tab:screening} reports the benchmark screening described in Section~\ref{sec:benchmarks}: for each of the eleven candidate benchmarks, 30 randomly drawn items were queried $K{=}10$ times by each of the three models, and an item counts as estimable when at least one model's observed accuracy falls in $[0.1,\,0.9]$. The screening run for ComplexBench did not complete.

\begin{table}[htbp]
\centering
\caption{Benchmark screening: estimable items out of 30 per candidate ($K{=}10$, three models). One benchmark per slot was selected, requiring a share of at least 35\% and taking the highest-share candidate per slot}
\label{tab:screening}
\small
\begin{tabular}{llrrl}
\toprule
Slot & Candidate & Estimable & Share (\%) & Outcome \\
\midrule
\multirow{4}{*}{Knowledge} & SuperGPQA & 13/30 & 43.3 & Selected \\
 & GPQA Diamond & 8/30 & 26.7 & Not selected \\
 & SuperGPQA hard & 7/30 & 23.3 & Not selected \\
 & MMLU-Pro & 2/30 & 6.7 & Not selected \\
\midrule
\multirow{3}{*}{Mathematics} & Omni-MATH hard & 21/30 & 70.0 & Selected \\
 & OlymMATH & 16/30 & 53.3 & Not selected \\
 & OlympiadBench text math & 11/30 & 36.7 & Not selected \\
\midrule
\multirow{4}{*}{Instruction following} & IFBench & 17/30 & 56.7 & Selected \\
 & ComplexBench & 11/30 & 36.7 & Incomplete \\
 & IFEval & 8/30 & 26.7 & Not selected \\
 & IFEval++ & 3/30 & 10.0 & Not selected \\
\bottomrule
\end{tabular}
\end{table}

\section{Single-draw comparison}
\label{app:single-draw}

For each item and model, one trial was drawn uniformly at random from the $K{=}50$ archived trials under a fixed seed, and the classification procedure of Section~\ref{sec:analysis} was applied to the resulting single observations. A comparison of two single binary observations never reaches significance under the Fisher/Benjamini--Hochberg criteria, so every item on every edge is classified as inconclusive and none of the 457 reliable item-level changes is recovered. Table~\ref{tab:single-draw} reports the raw correct-to-incorrect and incorrect-to-correct flips observed in the single draws; Table~\ref{tab:flip-cross} cross-tabulates these flips against the $K{=}50$ classification, aggregated over the nine migration--benchmark cells.

\begin{table}[htbp]
\centering
\caption{Single-draw comparison. Reliable changes under the full $K{=}50$ procedure, reliable changes detected from single draws, and raw flips between the single draws of source and target model (I$\to$C: incorrect to correct; C$\to$I: correct to incorrect)}
\label{tab:single-draw}
\small
\begin{tabular}{ll rrrrr}
\toprule
Edge & Benchmark & Items & $K{=}50$ rel.\ changes & Single-draw rel.\ changes & I$\to$C & C$\to$I \\
\midrule
\multirow{3}{*}{5.4$\to$5.5} & SuperGPQA & 500 & 66 & 0 & 27 & 20 \\
 & Omni-MATH hard & 100 & 23 & 0 & 14 & 4 \\
 & IFBench & 300 & 59 & 0 & 42 & 19 \\
\midrule
\multirow{3}{*}{5.5$\to$Sol} & SuperGPQA & 500 & 60 & 0 & 28 & 19 \\
 & Omni-MATH hard & 100 & 24 & 0 & 6 & 10 \\
 & IFBench & 300 & 60 & 0 & 25 & 47 \\
\midrule
\multirow{3}{*}{5.4$\to$Sol} & SuperGPQA & 500 & 78 & 0 & 38 & 22 \\
 & Omni-MATH hard & 100 & 15 & 0 & 9 & 3 \\
 & IFBench & 300 & 72 & 0 & 32 & 31 \\
\bottomrule
\end{tabular}
\end{table}

\begin{table}[htbp]
\centering
\caption{Raw single-draw flips cross-tabulated against the $K{=}50$ classification, aggregated over the nine migration--benchmark cells (2{,}700 item--edge pairs)}
\label{tab:flip-cross}
\small
\begin{tabular}{l rrrr}
\toprule
$K{=}50$ classification & I$\to$C & C$\to$I & Unchanged & Total \\
\midrule
Reliably improved & 120 & 8 & 127 & 255 \\
Reliably regressed & 10 & 87 & 105 & 202 \\
Practically equivalent & 25 & 26 & 1817 & 1868 \\
Inconclusive & 66 & 54 & 255 & 375 \\
\midrule
Total & 221 & 175 & 2304 & 2700 \\
\bottomrule
\end{tabular}
\end{table}

\section{Sensitivity to the practical-significance threshold}
\label{app:epsilon}

Table~\ref{tab:epsilon} reports the reliable-improvement, reliable-regression, and reliable-change shares of each migration--benchmark cell when the practical-significance threshold varies over $\varepsilon \in \{0.10, 0.15, 0.20, 0.25\}$, holding the Fisher/Benjamini--Hochberg criterion fixed. The rows with $\varepsilon = 0.20$ correspond to Table~\ref{tab:main}.

\begin{table}[htbp]
\centering
\caption{$P^{+}$, $P^{-}$, and $P^{\mathit{chg}}$ per migration--benchmark cell for four values of the practical-significance threshold~$\varepsilon$. All shares in percent}
\label{tab:epsilon}
\renewcommand{\baselinestretch}{1}\footnotesize
\begin{tabular}{ll l rrr}
\toprule
Edge & Benchmark & $\varepsilon$ & $P^{+}$ & $P^{-}$ & $P^{\mathit{chg}}$ \\
\midrule
\multirow{12}{*}{5.4$\to$5.5} & \multirow{4}{*}{SuperGPQA} & 0.10 & 8.4 & 5.2 & 13.6 \\
 &  & 0.15 & 8.4 & 5.2 & 13.6 \\
 &  & 0.20 & 8.2 & 5.0 & 13.2 \\
 &  & 0.25 & 8.0 & 4.6 & 12.6 \\
\cmidrule{2-6}
 & \multirow{4}{*}{Omni-MATH hard} & 0.10 & 19.0 & 6.0 & 25.0 \\
 &  & 0.15 & 19.0 & 6.0 & 25.0 \\
 &  & 0.20 & 17.0 & 6.0 & 23.0 \\
 &  & 0.25 & 15.0 & 3.0 & 18.0 \\
\cmidrule{2-6}
 & \multirow{4}{*}{IFBench} & 0.10 & 12.7 & 9.0 & 21.7 \\
 &  & 0.15 & 12.7 & 9.0 & 21.7 \\
 &  & 0.20 & 11.3 & 8.3 & 19.7 \\
 &  & 0.25 & 10.0 & 7.3 & 17.3 \\
\midrule
\multirow{12}{*}{5.5$\to$Sol} & \multirow{4}{*}{SuperGPQA} & 0.10 & 8.0 & 4.4 & 12.4 \\
 &  & 0.15 & 8.0 & 4.4 & 12.4 \\
 &  & 0.20 & 7.6 & 4.4 & 12.0 \\
 &  & 0.25 & 6.6 & 3.8 & 10.4 \\
\cmidrule{2-6}
 & \multirow{4}{*}{Omni-MATH hard} & 0.10 & 10.0 & 16.0 & 26.0 \\
 &  & 0.15 & 10.0 & 16.0 & 26.0 \\
 &  & 0.20 & 9.0 & 15.0 & 24.0 \\
 &  & 0.25 & 6.0 & 12.0 & 18.0 \\
\cmidrule{2-6}
 & \multirow{4}{*}{IFBench} & 0.10 & 7.0 & 14.0 & 21.0 \\
 &  & 0.15 & 7.0 & 14.0 & 21.0 \\
 &  & 0.20 & 6.7 & 13.3 & 20.0 \\
 &  & 0.25 & 6.0 & 12.7 & 18.7 \\
\midrule
\multirow{12}{*}{5.4$\to$Sol} & \multirow{4}{*}{SuperGPQA} & 0.10 & 11.4 & 5.0 & 16.4 \\
 &  & 0.15 & 11.4 & 5.0 & 16.4 \\
 &  & 0.20 & 10.8 & 4.8 & 15.6 \\
 &  & 0.25 & 10.2 & 4.6 & 14.8 \\
\cmidrule{2-6}
 & \multirow{4}{*}{Omni-MATH hard} & 0.10 & 11.0 & 5.0 & 16.0 \\
 &  & 0.15 & 11.0 & 5.0 & 16.0 \\
 &  & 0.20 & 10.0 & 5.0 & 15.0 \\
 &  & 0.25 & 8.0 & 3.0 & 11.0 \\
\cmidrule{2-6}
 & \multirow{4}{*}{IFBench} & 0.10 & 11.7 & 14.0 & 25.7 \\
 &  & 0.15 & 11.3 & 14.0 & 25.3 \\
 &  & 0.20 & 10.7 & 13.3 & 24.0 \\
 &  & 0.25 & 9.3 & 12.3 & 21.7 \\
\bottomrule
\end{tabular}
\end{table}

\section{Token consumption}
\label{app:tokens}

Table~\ref{tab:tokens} summarises per-request token counts for each benchmark and model over the 135{,}000 archived responses; Table~\ref{tab:token-edges} reports the change in mean total tokens per request on each migration edge.

\begin{table}[htbp]
\centering
\caption{Per-request token counts by benchmark and model. Mean input and output tokens, and mean, median, 5th and 95th percentile of total tokens}
\label{tab:tokens}
\small
\begin{tabular}{ll rrrrrr}
\toprule
 & & \multicolumn{2}{c}{Mean} & \multicolumn{4}{c}{Total tokens} \\
\cmidrule(lr){3-4}\cmidrule(lr){5-8}
Benchmark & Model & Input & Output & Mean & Median & P5 & P95 \\
\midrule
\multirow{3}{*}{SuperGPQA} & GPT-5.4 & 543.2 & 555.2 & 1098.4 & 758 & 468 & 2692 \\
 & GPT-5.5 & 543.2 & 543.5 & 1086.7 & 824 & 494 & 2767 \\
 & Sol & 543.8 & 409.7 & 953.5 & 702 & 476 & 2244 \\
\midrule
\multirow{3}{*}{Omni-MATH hard} & GPT-5.4 & 444.8 & 7653.3 & 8098.1 & 5732 & 1521 & 20779 \\
 & GPT-5.5 & 444.8 & 3705.3 & 4150.1 & 3684 & 991 & 8751 \\
 & Sol & 444.8 & 3766.1 & 4211.0 & 3527 & 1139 & 9128 \\
\midrule
\multirow{3}{*}{IFBench} & GPT-5.4 & 375.1 & 1030.2 & 1405.3 & 997 & 461 & 3699 \\
 & GPT-5.5 & 375.3 & 981.8 & 1357.1 & 992 & 466 & 3339 \\
 & Sol & 394.1 & 1452.9 & 1847.0 & 1118 & 443 & 5613 \\
\bottomrule
\end{tabular}
\end{table}

\begin{table}[htbp]
\centering
\caption{Mean total tokens per request on each migration edge}
\label{tab:token-edges}
\small
\begin{tabular}{ll rrr}
\toprule
Edge & Benchmark & Source & Target & Change \\
\midrule
\multirow{3}{*}{5.4$\to$5.5} & SuperGPQA & 1098.4 & 1086.7 & $-$11.7 \\
 & Omni-MATH hard & 8098.1 & 4150.1 & $-$3948.0 \\
 & IFBench & 1405.3 & 1357.1 & $-$48.1 \\
\midrule
\multirow{3}{*}{5.5$\to$Sol} & SuperGPQA & 1086.7 & 953.5 & $-$133.2 \\
 & Omni-MATH hard & 4150.1 & 4211.0 & $+$60.9 \\
 & IFBench & 1357.1 & 1847.0 & $+$489.8 \\
\midrule
\multirow{3}{*}{5.4$\to$Sol} & SuperGPQA & 1098.4 & 953.5 & $-$144.9 \\
 & Omni-MATH hard & 8098.1 & 4211.0 & $-$3887.2 \\
 & IFBench & 1405.3 & 1847.0 & $+$441.7 \\
\bottomrule
\end{tabular}
\end{table}

\section{Response category breakdown}
\label{app:categories}

Table~\ref{tab:categories} reports the six-category classification of Section~\ref{sec:scoring} for all archived responses. Truncation and API error do not occur in the final response matrix; refusals occur only on IFBench.

\begin{table}[htbp]
\centering
\caption{Six-category response classification by benchmark and model (counts; $N$ is items~$\times$~$K{=}50$)}
\label{tab:categories}
\small
\begin{tabular}{ll rrrrrrr}
\toprule
Benchmark & Model & $N$ & Correct & Wrong & Format & Refusal & Trunc. & API err. \\
\midrule
\multirow{3}{*}{SuperGPQA} & GPT-5.4 & 25000 & 15841 & 8913 & 246 & 0 & 0 & 0 \\
 & GPT-5.5 & 25000 & 16424 & 8205 & 371 & 0 & 0 & 0 \\
 & Sol & 25000 & 16780 & 7932 & 288 & 0 & 0 & 0 \\
\midrule
\multirow{3}{*}{Omni-MATH hard} & GPT-5.4 & 5000 & 2285 & 2689 & 26 & 0 & 0 & 0 \\
 & GPT-5.5 & 5000 & 2648 & 2326 & 26 & 0 & 0 & 0 \\
 & Sol & 5000 & 2445 & 2555 & 0 & 0 & 0 & 0 \\
\midrule
\multirow{3}{*}{IFBench} & GPT-5.4 & 15000 & 9418 & 4635 & 945 & 2 & 0 & 0 \\
 & GPT-5.5 & 15000 & 9696 & 4528 & 771 & 5 & 0 & 0 \\
 & Sol & 15000 & 9111 & 4534 & 1352 & 3 & 0 & 0 \\
\bottomrule
\end{tabular}
\end{table}

\section{Difficulty stratification}
\label{app:difficulty}

Table~\ref{tab:difficulty} reports mean strict accuracy per model and the per-edge change within strata defined by benchmark-native metadata: SuperGPQA provides difficulty and discipline labels, and Omni-MATH hard provides a difficulty rating. IFBench provides no comparable difficulty metadata and is therefore not stratified.

\begin{table}[htbp]
\centering
\caption{Mean strict accuracy (\%) per model and change per edge (percentage points) within benchmark-native strata}
\label{tab:difficulty}
\small
\begin{tabular}{ll r rrr rrr}
\toprule
Benchmark & Stratum & Items & 5.4 & 5.5 & Sol & $\Delta_{5.4\to5.5}$ & $\Delta_{5.5\to\mathrm{Sol}}$ & $\Delta_{5.4\to\mathrm{Sol}}$ \\
\midrule
\multirow{16}{*}{SuperGPQA} & Difficulty easy & 156 & 64.0 & 69.3 & 72.4 & $+$5.3 & $+$3.1 & $+$8.4 \\
 & Difficulty middle & 204 & 69.8 & 70.8 & 72.6 & $+$0.9 & $+$1.8 & $+$2.7 \\
 & Difficulty hard & 140 & 53.2 & 54.3 & 53.3 & $+$1.0 & $-$1.0 & $+$0.1 \\
 & Agronomy & 10 & 23.2 & 23.4 & 27.4 & $+$0.2 & $+$4.0 & $+$4.2 \\
 & Economics & 17 & 74.0 & 76.7 & 78.6 & $+$2.7 & $+$1.9 & $+$4.6 \\
 & Education & 10 & 50.6 & 67.2 & 72.8 & $+$16.6 & $+$5.6 & $+$22.2 \\
 & Engineering & 135 & 66.5 & 70.3 & 70.3 & $+$3.8 & $-$0.0 & $+$3.7 \\
 & History & 16 & 65.1 & 84.2 & 82.9 & $+$19.1 & $-$1.4 & $+$17.8 \\
 & Law & 6 & 37.3 & 61.3 & 55.3 & $+$24.0 & $-$6.0 & $+$18.0 \\
 & Literature and Arts & 41 & 55.9 & 59.9 & 64.0 & $+$4.0 & $+$4.2 & $+$8.2 \\
 & Management & 12 & 57.2 & 57.2 & 64.0 & 0.0 & $+$6.8 & $+$6.8 \\
 & Medicine & 57 & 71.3 & 68.1 & 74.4 & $-$3.2 & $+$6.2 & $+$3.0 \\
 & Military Science & 4 & 54.5 & 27.0 & 43.0 & $-$27.5 & $+$16.0 & $-$11.5 \\
 & Philosophy & 11 & 71.1 & 72.0 & 78.0 & $+$0.9 & $+$6.0 & $+$6.9 \\
 & Science & 180 & 62.8 & 63.4 & 62.9 & $+$0.6 & $-$0.5 & $+$0.1 \\
 & Sociology & 1 & 100.0 & 100.0 & 100.0 & 0.0 & 0.0 & 0.0 \\
\midrule
\multirow{4}{*}{Omni-MATH hard} & Difficulty 8.0 & 61 & 43.8 & 52.0 & 47.6 & $+$8.2 & $-$4.4 & $+$3.8 \\
 & Difficulty 8.5 & 2 & 49.0 & 50.0 & 49.0 & $+$1.0 & $-$1.0 & 0.0 \\
 & Difficulty 9.0 & 32 & 50.4 & 53.8 & 52.9 & $+$3.4 & $-$0.9 & $+$2.5 \\
 & Difficulty 9.5 & 5 & 37.2 & 60.4 & 39.2 & $+$23.2 & $-$21.2 & $+$2.0 \\
\bottomrule
\end{tabular}
\end{table}

\end{document}